\magnification=1200
\overfullrule=0pt
\baselineskip24pt


\font\tenmsb=msbm10
\font\sevenmsb=msbm7
\font\fivemsb=msbm5
\newfam\msbfam

\textfont\msbfam=\tenmsb 
\scriptfont\msbfam=\sevenmsb
\scriptscriptfont\msbfam=\fivemsb

\catcode`\@=11
\def\Bbb{\ifmmode\let\next\Bbb@\else
 \def\next{\errmessage{Use \string\Bbb\space only in math mode}}\fi\next}
\def\Bbb@#1{{\Bbb@@{#1}}}
\def\Bbb@@#1{\fam\msbfam#1}
\catcode`\@=12

\def\hexnumber@#1{\ifcase#1 0\or 1\or 2\or 3\or 4\or 5\or 6\or 7\or 8\or
 9\or A\or B\or C\or D\or E\or F\fi}


\def\mno{\medskip\noindent}

\def\today{\noindent\number\day 
\space\ifcase\month
 \or January
 \or February
 \or March
 \or April
 \or May
 \or June
 \or July
 \or August
 \or September
 \or October
 \or November
 \or December
\fi \space\number\year}


 
\def\Z  {{\Bbb Z}} 




\def\b {\beta} 
\def\g {\gamma}

\def\s {\sigma}


\def\qed{\vbox{\hrule width.5em height1.1ex depth.2ex}}

\medskip
\centerline {\bf IMPROVED PEIERLS ARGUMENT FOR HIGH DIMENSIONAL}
\centerline {\bf ISING MODELS}

\bigskip
\bigskip
\centerline{ 
{\bf J.L. Lebowitz $^{1,2}$ 
\footnote{ }{$^1$ \rm Department of Mathematics and Physics, Rutgers
University, New Brunswick, N.J. 08903 USA}
\footnote{ }{$^2$ \rm Work supported in part by NSF
grant DMR 92-13424} 
and ~~~~~A.E.Mazel $^{1,2,3}$
\footnote{ }{$^3$ \rm International Institute of Earthquake
prediction Theory and Theoretical Geophysics, 113556 Mos\-cow, Russia} }}

\bigskip\noindent
{\bf Abstract.} {We consider the low temperature expansion for the Ising
model on $\Z^d$, $d \ge 2$, with ferromagnetic nearest neighbor
interactions in terms of Peierls contours.  We prove that the expansion
converges for all temperatures smaller than $C d (\log d)^{-1}$, which is
the correct order in $d$.}


\bigskip
\mno
{\bf 1. Introduction  and Result}

\medskip 
We consider the Ising model without an external magnetic field on the
$d$-dimensional cubic lattice $\Z^d$. The model is defined by the formal
Hamiltonian
$$H(\s)=-\sum_{(x,y)} \s_x \s_y, \eqno{(1.1)}$$
where the configuration $\s \in \{ -1, +1\}^{\Z^d}$ takes values $\s_x$
and $\s_y$ at sites $x$ and $y$ of $\Z^d$ and the sum is taken over all
nearest neighbor bonds $(x,y)$ of $\Z^d$. It is known [BKLS] that the
critical inverse temperature $\b_{cr}$ of the Ising model behaves like $1
\over 4d$ for large $d$. An estimate $\b_{cr} > {C_1/d}$ for
some $C_1 \le {1 \over 4} $ can be easily derived from the standard high 
temperature expansions [R] or by other arguments [G], [S]. Here and below 
we denote by $C_1$, $C_2,
\ldots$  positive absolute constants. 

At first sight it is surprising that the best currently available
upper bound, $\b_{p} < C_2$, given by the Peierls argument [R], is
very far from the true value of $\b_{cr}$ when $d$ is large.  Naively
one would like to have from the low temperature expansion an upper
bound of the form $\b_{p} < C_3/d$, i.e. at least of the same order as
the lower bound.  Unfortunately this is impossible. The Peierls
argument, i.e. the convergence of the low temperature expansion,
automatically implies the absence of percolation of the minority phase
while it is known [ABL] that for the inverse temperature $\b$ higher
than $\b_{cr}$ but lower than $C_4 \log d/d$ the minority phase does
percolate. Thus for large $d$ the upper estimate for $\b_{cr}$ given
by the Peierls argument can not be close to $\b_{cr}$. Nevertheless it
is interesting to understand what is the radius of convergence for the
low temperature expansion written in terms of Peierls contours. The
answer, which is correct up to the constant factor, is given by

\mno
{\bf Theorem 1.1} {\sl For the Ising model (1.1) the low temperature
expansion, written in terms of Peierls contours, is convergent for $\b \ge
64 (\log d)/d$.}

\medskip
The geometric problem closely related to this theorem is the upper
bound for the number, $\sharp (n)$, of Peierls contours of size
$n$. The best previous estimate was $\sharp (n) \le 3^n$
[R]. Now we improve this estimate

\mno
{\bf Corollary 1.2} {\sl The number, $\sharp (n)$, of different Peierls
contours of size $n$ is less than $\exp [64n (\log d)/d]$.}

\medskip
On the other hand it is not hard to see that 

\mno
{\bf Lemma 1.3} {\sl The number $\sharp (n)$ is larger than
$\exp[(n-2d)(\log d)/(2d-2)]$.}

\mno
{\bf Proof.} Consider a chain of $k$ lattice sites which starts at a 
given $x \in \Z^d$ and every next site is obtained from the previous one
by the unit shift in one of the positive coordinate directions. Clearly
one has $d^{k-1}$ different chains of that type. Take the contour which is
the boundary of the union of the unit $d$-dimensional cubes centered at
the sites of the chain. The size of this contour is $(2d-2)k +2$ and
different chains produce different contours. \qed

\medskip
Thus our estimate is correct in order though the constant $64$ is
certainly too large.

\bigskip
\mno
{\bf 2. Proof of Theorem}

\medskip 
We begin with some geometric notions which we need to define Peierls
contours. A {\it plaquette} is a unit $(d-1)$-dimensional cube from the
dual lattice centered at the middle of some bond of the initial lattice. Two
plaquettes are called adjacent if they have a common $(d-2)$-dimensional
face. Two lattice sites are called adjacent if they are endpoints of some
lattice bond. A plaquette and a lattice site are adjacent if this
plaquette intersects one of the lattice bonds incident on this site.  A
lattice site and a $(d-2)$-dimensional face are adjacent if the site is
adjacent to one of four plaquettes incident to this face.

A set of plaquettes is connected if any two its plaquettes belong to a
chain of pairwise adjacent plaquettes from this set. Similarly a lattice
subset is connected if any two of its sites belong to a chain of pairwise
adjacent sites from this set.

Consider a configuration $\s$ containing a finite number of sites $x$ at
which $\sigma_x = -1$.  
For every unit lattice bond $(x,y)$ having $\s_x \s_y =-1$ draw a
plaquette orthogonal to $(x,y)$. Such plaquettes form a closed surface
consisting of several connected components which are known as Peierls
contours. Clearly a {\it contour} $\g$ is just a connected closed
plaquette surface separating a finite set $\bar \g \subset \Z^d$ called
the interior of $\g$ from its complement $\bar \g^c = \Z^d \setminus \bar
\g$ called the exterior of $\g$.

Two contours are called {\it compatible} if their union is not a connected
set of plaquettes. A collection of contours is called compatible if any
two contours from this collection are compatible. It is not hard to see
that the correspondence between finite compatible collections of contours
and configurations $\s$ with a finite number of ``$-$'' spins is
one-to-one.

The convergence of the low temperature expansion means the existence of
an absolutely convergent polymer series for the logarithm of the
partition function in any finite region with ``$+$'' (and by symmetry
``$-$'') boundary conditions. This series is the sum of statistical
weights of so called polymers belonging to the region. Every
contributing polymer is a finite family of contours which can be indexed
in such a way that every next contour is incompatible with at least one
of previous contours. The notion of polymer is dual to that of the
compatible collection of contours. The statistical weight of the polymer
is the product of the statistical weights of contributing contours times
a combinatorial factor of more complicated structure. The details and
precise definitions can be found in any reference on cluster (polymer)
expansions. In particular [KP], [MS] or [D] contain general theorems
which, applied to the Ising model, give us

\mno
{\bf Lemma 2.1} {\sl The polymer expansion constructed for the Ising
model in terms of Peierls contours is convergent at inverse temperature
$\b$ if there exists a positive function $a(\g)$ such that for any
contour $\g$
$$\sum_{\g'} w(\g') e^{a(\g')} \le a(\g), \eqno{(2.1)}$$
where the sum is taken over all contours $\g'$ incompatible with
$\g$, $w(\g')= \exp ( -\b |\g'| )$ is the statistical weight of contour
$\g'$ and $|\g'|$ denotes the number of plaquettes in $\g'$. }

\medskip
According to the Peierls argument [P] the probability that $\s_0=-1$ in the
Gibbs state $\langle \; \cdot\; \rangle^+$ with ``$+$'' boundary
condition is equal to the probability that an odd number of contours
surround the origin. This is less than the probability that $\bar \g \ni
0$ for at least one contour $\g$. A given contour $\g$ has a probability
not exceeding $w(\g)$. Hence the magnetization $\langle \s_0
\rangle^+$ is positive as soon as
$$\sum_{\g:\; \bar \g \ni 0} w(\g)  < {1 \over 2}. \eqno{(2.2)}$$

Condition (2.1) is stronger than (2.2) but has a similar nature. To
clarify the difference observe that $\g'$ is incompatible with $\g$ iff
it contains at least one of the $(d-2)$-dimensional faces of $\g$.
Therefore $\bar \g'$ contains one of the lattice sites adjacent to some
$(d-2)$-dimensional face of $\g$. The number of $(d-2)$-dimensional
faces in $\g$ does not exceed $(d-1)|\g|$ as such a face is shared by
two or four plaquettes from $\g$. Hence the number of adjacent lattice
sites is less than $4d|\g|$ and (2.1) is satisfied for $a(\g)=\b
\exp[-d\beta/4]|\g|$ if 
$$\sum_{\g:\; \bar \g \ni x} w(\g) \exp \left( \b e^{-{d\b \over 4}}
|\g| \right) \le {\b \over 4d} e^{-{d\b \over 4}}, \eqno{(2.3)}$$ 
where the sum is taken over all contours surrounding a site $x \in
\Z^d$. 

Estimate (2.3) is stronger than (2.2) because $\exp \left( \b e^{-{d\b
\over 4}} |\g| \right)>1$ and ${\b \over 4d} e^{-{d\b \over 4}} \le
e^{-1}d^{-2}< 1/2$. In the proof of Lemma~2.1 the factor $\exp \left( \b
e^{-{d\b \over 4}} |\g| \right)$ is used to dominate the combinatorial
factor contributing to the statistical weight of the polymer. The
control obtained is enough to see that the sum of the absolute values of
the statistical weights of all polymers surrounding a site $x \in \Z^d$
is smaller than ${\b \over 4d} e^{-{d\b \over 4}}$. This gives the
convergence of the polymer expansion implying various nice properties,
e.g. an exponential decay of correlations in the state $\langle \;
\cdot\; \rangle^+$.

In the rest of the paper we check that (2.3) is true for $\b \ge 64
(\log d)/d$. A contour $\g$ is called {\it primitive} if it can not be
partitioned into two contours $\g'$ and $\g''$. If $\g$ is not primitive
then the corresponding $\g'$ and $\g''$ have no common plaquettes but
have common $(d-2)$-dimensional faces. In particular
$w(\g)=w(\g')w(\g'')$. For $\g$ surrounding a given site $x$ consider
some decomposition of $\g$ into primitive subcontours: note that such a
decomposition may not be unique. For a fixed decomposition the set of
corresponding primitive subcontours can be naturally provided with a
tree-like structure. The root of the tree is the primitive subcontour
$\g_0$ surrounding $x$. The fist-level subcontours $\g_{1,i_1}$ are the
subcontours having a common $(d-2)$-dimensional face with $\g_0$. The
second-level subcontours $\g_{2,i_2}$ are the subcontours (not included
in the previous levels) which have a common $(d-2)$-dimensional face
with at least one of the first-level subcontours. Generally the
subcontours of the $n$-th level $\g_{n,i_n}$ are the subcontours which
have a common $(d-2)$-dimensional face with at least one of the
subcontours from level $n-1$ and are not included in $\cup_{k=0}^{n-1}
\cup_{i_k} \g_{k, i_k}$.

We will now show that (2.3) will be satisfied when the following inequality
holds: 
$$\sum_{\g:\; \bar \g \ni x,\; \g \; {\rm is}\;{\rm primitive}} w(\g) \exp
\left( 2\b e^{-{d\b \over 4}} |\g| \right) \le {\b \over 4d} e^{-{d\b
\over 4}}, \eqno{(2.4)}$$
with the sum over primitive contours only.

Denote by $n(\g)$ the number of levels in the decomposition of $\g$ into
primitive subcontours. By induction in $n(\g)$ we check that, for any
$n$, (2.4) implies
$$
\sum_{\g:\; \bar \g \ni x, n(\g)\le n} w(\g) \exp \left( \b e^{-{d\b
\over 4}} |\g| \right) \le {\b \over 4d} e^{-{d\b \over 4}},
\eqno{(2.5)}
$$ 
and hence (2.3). For $n=1$ (2.5) clearly follows from
(2.4). Suppose now that (2.5) is true for $n=N-1$ and consider the case
$n=N$. Without $\g_0$ the subcontours from $\cup_{k=1}^{N} \cup_{i_k}
\g_{k, i_k}$ are decoupled into subtrees with $\g_{1,i_1}$ serving as new
roots. Clearly for each subtree the number of levels does not exceed
$N-1$. By construction the subcontour $\g_{1,i_1}$ surrounds a site $y$
adjacent to some $(d-2)$-dimensional face of $\g_0$ and the set $A(\g_0)$
of all such sites has the cardinality $|A(\g_0)| \le 4d|\g_0|$. Using the
induction hypothesis (in the second inequality below) one obtains
$$
\eqalignno{
&\sum_{\g:\; \bar \g \ni x, n(\g)\le N} w(\g) \exp \left( \b e^{-{d\b
\over 4}} |\g| \right) \cr
\le &\sum_{\g_0:\;  \bar \g_0 \ni x} w(\g_0) \exp \left( \b e^{-{d\b \over
4}} |\g_0| \right) \prod_{y \in A(\g_0)} \left( 1+  \sum_{\g:\; \bar \g
\ni y, n(\g) \le N-1} w(\g) \exp \left( \b e^{-{d\b \over 4}} |\g| \right)
\right) \cr
\le &\sum_{\g_0:\;  \bar \g_0 \ni x} w(\g_0) \exp \left( \b e^{-{d\b \over
4}} |\g_0| \right) \prod_{y \in A(\g_0)} \left( 1+ {\b \over 4d} e^{-{d\b
\over 4}} \right) \cr
\le &\sum_{\g_0:\;  \bar \g_0 \ni x} w(\g_0) \exp \left( \b e^{-{d\b \over
4}} |\g_0| \right) \exp \left(4d|\g_0| {\b \over 4d} e^{-{d\b \over 4}}
\right) \cr
= &\sum_{\g_0:\;  \bar \g_0 \ni x} w(\g_0) \exp \left( 2\b e^{-{d\b \over
4}} |\g_0| \right) \cr
\le &{\b \over 4d} e^{-{d\b \over 4}}, &(2.6)\cr}
$$ 
which reproduces (2.5) for $n=N$.

{From} now on we discuss primitive contours only and we call them  contours.
Denote by $\g_i$, $i=1, \ldots, d$ the set of plaquettes of $\g$
orthogonal to the coordinate axis number $i$. Let $\displaystyle
|\g_{i^*}|= \min_i |\g_i|$. The direction $i^*$ is called $\g$-vertical
and all plaquettes of $\g$ are separated into horizontal ones, i.e. those
belonging to $\gamma_i$, 
$\g_{i^*}$, and vertical ones, i.e. those from $\g \setminus
\g_{i^*}$. {}From now on we denote them $\g^{hor}$ and $\g^{ver}$ respectively.

Consider a site $x \in \Z^d$ and a contour $\g$ surrounding $x$. Draw a
$\g$-vertical line through $x$. This line intersects some  plaquette $p
\in \g^{hor}$ and the distance from $x$ to $p$ is less than $|\g^{ver}|
\over 2d-2$. Now it is clear that the sum in (2.4) does not exceed
$$\sum_{\g:\; \g^{hor} \ni p} d {|\g^{ver}| \over 2d-2} w(\g) \exp \left(
2\b e^{-{d\b \over 4}} |\g| \right), \eqno{(2.7)}$$ 
where $p$ is fixed and the factor $d$ counts the number of choices for the
vertical direction.

Our main observation is the following simple estimate
$$
|\g| \ge {d \over 2}|\g^{hor}| + {1 \over 2} |\g^{ver}|, \eqno{(2.8)}
$$
which is an immediate consequence of the definition of $\g^{hor}$. 
It reduces (2.7) to
$$\sum_{\g:\; \g^{hor} \ni p} |\g^{ver}|\exp \left( 2\b e^{-{d\b \over
4}} |\g| \right) \exp \left( -{d \b \over 2} |\g^{hor}| -{\b \over 2}
|\g^{ver}| \right) \le {\b \over 4d} e^{-{d\b \over 4}}. \eqno{(2.9)}$$
The elementary inequalities:
$$2\b e^{-{d\b \over 4}} \le {\b \over 16} \quad {\rm for}\quad \b \ge
{20 \log2 \over d}, \eqno{(2.10)}$$
$$|\g^{ver}| e^{-{7\b \over 16} |\g^{ver}|} \le  e^{-{6\b \over 16}
|\g^{ver}|} \quad {\rm for}\quad \b \ge {16 \log |\g^{ver}| \over 
|\g^{ver}|}, \eqno{(2.11)}$$
$${16 \log d \over d} \ge {16 \log (2d-2) \over 2d-2 } \ge {16 \log
|\g^{ver}| \over |\g^{ver}|} \quad {\rm for}\quad d \ge 3 \eqno{(2.12)}$$
reduce (2.9) to the bound
$$\sum_{\g:\; \g^{hor} \ni p} \exp \left( -{3d \b \over 8} |\g^{hor}|
-{3\b \over 8} |\g^{ver}| \right) \le {\b \over 4d} e^{-{d\b \over 4}}
\eqno{(2.13)}$$

Let a {\it floor}  of a contour $\g$ be a connected component of $\g^{hor}$. 

\mno
{\bf Lemma 2.2} {\sl A contour $\g$ is uniquely defined by the family of
its floors.}

\mno 
{\bf Proof.} To show this we reconstruct all plaquettes of $\g$ from the
family of its floors. We start with the floors of $\g$ situated at the
minimal vertical level which we denote by $m$. Their boundary consists of
$(d-2)$-dimensional faces. Each of these faces has a unique vertical
plaquette growing from it in the positive vertical direction. We add all
these vertical plaquettes to the set of already reconstructed
plaquettes. All floors of $\g$ situated at vertical level $m+1$ are added
next. The new boundary set of already reconstructed plaquettes consists of
$(d-2)$-dimensional faces situated at level $m+1$. Again each of these
faces has a unique vertical plaquette growing from it in the positive
vertical direction. Adding all these vertical plaquettes one moves to the
level $m+2$. Then we add all floors situated at the level $m+2$ and so
on. The procedure terminates after a finite number of steps when the
boundary of the set of reconstructed plaquettes becomes empty. \qed

\medskip
The contour $\g$ is uniquely decomposed into horizontal floors
$\g^{hor}_i$ and vertical plaquette stacks $\g^{ver}_j$. A {\it vertical
plaquette stack} is a chain of pairwise adjacent vertical plaquettes
extending between two floors. None but the first and the last plaquettes
in the stack have a common $(d-2)$-dimensional face with the floors and
every next plaquette in the stack can be obtained from the previous
one by the unit vertical shift. One may associate with $\g$ an abstract
connected graph with floors being the vertices and vertical stacks being
the links of the graph. Note that in this graph two vertices may have more
than one link joining them. Let $T=\{\g^{hor}_0, \g^{hor}_{m,i_m},
\g^{ver}_{m,i_m} \}$ be a spanning tree of this graph. We assume that its
root, $\g^{hor}_0$, passes through the given plaquette $p$. The links of
the first level, $\g^{ver}_{1,i_1}$, join the root with the corresponding
vertices of the first level, $\g^{hor}_{1,i_1}$, and so on. We associate
with the tree $T$ the statistical weight
$$w(T)=\exp \left(-{3d\b \over 8} |\g^{hor}_0| - \sum_m \sum_{i_m} \left(
{3d\b \over 8} |\g^{hor}_{m,i_m}| + {3\b \over 8} |\g^{ver}_{m,i_m}|
\right) \right). \eqno{(2.14)}$$
Then in view of Lemma~2.2 bound (2.13) follows from
$$\sum_{T:\; \g^{hor}_0(T) \ni p} w(T) \le {\b \over 4d} e^{-{d\b \over 4}},
\eqno{(2.15)}$$
where the sum is taken over all abstract trees of the type described
above. In the same way as (2.3) follows from (2.4) one can
see that (2.15) is a consequence of
$$\sum_{(\g^{ver}, \g^{hor}):\; \g^{ver} \ni p} \exp \left(
-{3d\b \over 8} |\g^{hor}| -{3\b \over 8} |\g^{ver}| \right)
\exp \left( \b e^{-{d\b \over 4}} |\g^{hor}| \right) \le 
{\b \over 4d} e^{-{d\b \over 4}}. \eqno{(2.16)}$$
Here the sum is taken over all pairs $(\g^{ver}, \g^{hor})$ consisting of
the vertical stack, $\g^{ver}$, starting at the fixed plaquette $p$ and
the floor, $\g^{hor}$, connected to the end of the stack $\g^{ver}$. The
analogue of estimate (2.6) is applicable as the number of the
starting plaquettes for the first-level stacks growing from $\g^{hor}_0$
does not exceed $(2d-2)|\g^{hor}_0| \le 4d|\g^{hor}_0|$.

In view of (2.10) the bound (2.16), for $\b > {20 \log2 \over d}$, is
reduced to
$$\sum_{(\g^{ver}, \g^{hor}):\; \g^{ver} \ni p} \exp \left( -{5d\b \over
16} |\g^{hor}| -{5\b \over 16} |\g^{ver}| \right) \le  {\b \over 4d}
e^{-{d\b \over 4}}. \eqno{(2.17)}$$
Now we have
$$\eqalignno{
&\sum_{(\g^{ver}, \g^{hor}):\; \g^{ver} \ni p} \exp \left( -{5d\b \over
16} |\g^{hor}| -{5\b \over 16} |\g^{ver}| \right) \cr
\le &\sum_{|\g^{ver}|=1}^{\infty} \exp \left(-{5\b \over 16} |\g^{ver}|
\right) \sum_{|\g^{hor}|=1}^{\infty} \exp \left(-{5d\b \over 16}
|\g^{hor}| \right) (2d)^{|\g^{hor}|} \cr
= &{2d e^{-{5d\b \over 16}-{5\b \over 16}} \over \left( 1- 2d e^{-{5d\b
\over 16}} \right) \left( 1- e^{-{5\b \over 16}} \right)} \cr
\le &{32d \over 5\b}{e^{-{5d\b \over 16}} \over 1- e^{-{5d\b \over 16}}} \cr
\le &{\b \over 4d} e^{-{d\b \over 4}},  &(2.18) \cr}$$
where the last inequality is true for $\b \ge 64 {\log d \over d}$. \qed

\bigskip
\mno
{\bf 1. References}

\medskip 
\item{~[ABL]} M.Aizenman, J.Bricmont and J.L.Lebowitz, ``Percolation of
the Minority Spins in High-Dimensional Ising Models'', {\it J. Stat.
Phys.} {\bf 49}, N3/4, 859-865 (1987).
\item{[BKLS]} J.Bricmont, H.Kesten, J.L.Lebowitz and R.H.Schonmann, ``A
Note on the Ising Model in High Dimensions'', {\it Comm. Math. Phys.}
{\bf 122}, 597-607 (1989).
\item{~~~[D]} R.L.Dobrushin, ``Estimates of Semi-invariants for the
Ising Model at Low Temperatures'', Topics in statistical and theoretical
physics, {\it Amer. Math. Soc. Transl. (2)}, {\bf V177}, 59--81 (1996).
\item{~~~[G]} R. B. Griffiths, ``Correlations in Ising Ferromagnets, III.  
A Mean-Field Bound for Binary Correlations'', {\it Comm. Math. Phys.} {\bf
6}, 121-127 (1967).
\item{~~[KP]} R.Kotecky and D.Preiss, ``Cluster Expansion for Abstract
Polymer Models'', {\it Comm. Math. Phys.}, {\bf 103}, 491-498 (1986)
\item{~~[MS]} A.E.Mazel and Yu.M.Suhov, ``Ground States of Boson Quantum
Lattice Model'', {\it Amer. Math. Soc. Transl. (2)}, {\bf V171}, 185-226
(1996). 
\item{~~~[P]} R.Peierls, ``On Ising's Model of Ferromagnetism'', {\it
Proc. Camb. Phil. Soc.} {\bf 32}, 477-481 (1936).
\item{~~~[R]} D.Ruelle, ``Statistical Mechanics. Rigorous Results'', New
York: Benjamin, (1969).
\item{~~~[S]} B.Simon, ``A remark on Dobrushin's Uniqueness Theorem'', 
{\it Comm. Math. Phys.} {\bf 68}, 183-185 (1979).

\end